\documentstyle[floats,aps,prl,epsfig]{revtex}

\begin{document}
\topmargin -1.50cm \draft
\twocolumn[\hsize\textwidth\columnwidth\hsize\csname@twocolumnfalse%
\endcsname
\title{The Evolution of Quasiparticle Charge\\
in the Fractional Quantum Hall Regime}

\author{T. G. Griffiths, E. Comforti, M. Heiblum, Ady Stern and V. Umansky}
\address{Braun Center for Submicron Research, Dept. of Condensed Matter
Physics\\ Weizmann Institute of Science, Rehovot, Israel 76100}
\maketitle
\begin{abstract}
\noindent The charge of quasiparticles in a fractional quantum Hall (FQH) liquid, tunneling through a partly
reflecting constriction with transmission $t$, was determined via shot noise measurements. In the $\nu=1/3$ FQH
state, a charge smoothly evolving from $e^*=e/3$ for $t_{1/3}\cong1$ to $~e^*=e$ for $t_{1/3}\ll1$ was determined,
agreeing with chiral Luttinger liquid theory. In the $\nu=2/5$ FQH state the quasiparticle charge evolves smoothly
from $e^*=e/5$ at $t_{2/5}\cong1$ to a maximum charge less than $e^*=e/3$ at $t_{2/5}\ll1$. Thus it appears that
quasiparticles with an approximate charge $e/5$ pass a barrier they see as almost opaque.\\

\end{abstract}
\pacs{PACS numbers: 73.20.Hm, 71.10.Pm, 73.50.Td}]

\noindent The fractional quantum Hall (FQH) effect is a manifestation of the prominent and unique effects
resulting from the Coulomb interactions between electrons in a two-dimensional electron gas (2DEG) under the
influence of a strong magnetic field \cite{Prange87}. In this regime the lowest Landau level is partially
populated. Laughlin's seminal explanation of the FQH effect \cite{Laughlin83} involved the emergence of intriguing
fractionally charged quasiparticles. Recently, shot noise measurements confirmed the existence of such
quasiparticles with charge $e/3$ and $e/5$ at filling factors $\nu=1/3$ \cite{rafi97} and $\nu=2/5$
\cite{Reznikov99}, respectively. These experiments relied on the fact that shot noise, resulting from the granular
nature of the quasiparticles, is proportional to their charge. Since current flowing in an ideal Hall state is
noiseless \cite{Reznikov99} a quantum point contact (QPC) constriction was used to weakly reflect the incoming
current, leading to partitioning of the incoming carriers and hence to shot noise. A charge $e^*$ was then deduced
from the shot noise expression derived for non-interacting particles \cite{Lesovik89}.  In this paper, we extend
the range of QPC reflection to the strong back-scattering limit, where the apparent noise-producing quasiparticle
charge is expected to be different. Specifically, an opaque barrier is expected to allow only the tunneling of
electrons, as both sides of the barrier should be quantized in units of the electronic charge. How this charge
evolves is an important question in the understanding of the behavior of quasiparticles, and here we explore the
evolution of the charge of the $e/3$ and $e/5$ quasiparticles. We first briefly describe the expected dependence
of shot noise on charge and transmission.
\\
\\
At zero temperature $(T=0)$, the shot noise contribution of the $p$'th channel is \cite{Lesovik89,Reznikov98}:
\begin{equation}\label{schky}
S_{T=0}=2e^*Vg_pt_p(1-t_p),
\end{equation}
where $S$ is the low frequency $(f<<eV/h)$ spectral density of current fluctuations $(S\Delta f=\langle
i^2\rangle)$, $V$ the applied source-drain voltage, $g_p$ the conductance of the fully transmitted $p$'th channel
in the QPC, and $t_p$ is its transmission coefficient. This reduces to the well known {\em classical} Poissonian
expression for shot noise when $t_p\ll1$ (the 'Schottky equation'), $S_{T=0}=2eI$, with $I=Vg_pt_p$ the DC current
in the QPC.
\\
\\
The justification for the use of Eq.~(\ref{schky}) comes from current theoretical studies of shot noise in the FQH
regime, based on the chiral Luttinger liquid model.  They are applicable only for Laughlin's fractional states,
 $\nu=1/3$, $1/5$, etc. \cite{Kane94,Fendley95,Chamon95} (where the edge is composed of one channel only) and
not for more general filling factors. They predict the following:%
\begin{eqnarray}\label{limits}
S_{T=0}&=&2e^{*}Vg_{p}(1-t_{p})=2e^{*}I_{r}~~\hspace{1.25cm};t_{p}\approx 1,\nonumber
\\
S_{T=0}&=&2eVg_{p}t_{p}=2eI_{t}~~\hspace{2.5cm};t_{p}\approx 0,
\end{eqnarray}
where $I_r$ and $I_t$ are the reflected and transmitted DC currents, respectively.  The most important result of
Eq.~(\ref{limits}) is that the tunneling of quasiparticles with charge $e/3$, $e/5$, etc. in Laughlin states, at
weak reflection $(t_p\approx1)$, changes to that of electrons at strong reflection $(t_p\approx0)$.
\\
\\
One can gain insight into the characteristics of the expected shot noise in the FQH regime \cite{Reznikov99}, and
some insight into Eq.~(\ref{schky}), by considering the Composite Fermion (CF) model \cite{Jain89}.  In the
simplest approximation for the CF model the fractionally filled electronic Landau level with $\nu=p/(2p+1)$ is
identified as $p$ filled Landau levels of CFs, $\nu_{CF}=p$, with each CF consisting of an electron with {\em two}
attached magnetic flux quanta $\phi_0=h/e$. The effective magnetic field sensed by the CFs is $B-2n_sh/e$, with
$n_s$ the density of the 2DEG. Under this weaker effective magnetic field the CFs are approximated as weakly
interacting quasiparticles, flowing in separate and non-interacting edge channels, hence justifying the
application of the above-mentioned formulae for the noise. When the QPC constriction is reduced in width and the
conductance is in a transition between two different FQH plateaus of the series $p/(2p+1)$ only one edge channel
is partitioned. The others can be approximated as being perfectly transmitted.  Consequently, in Eqs.
(\ref{schky}) and (\ref{limits}), $p$ designates the CF edge channel that is being partitioned.  As examples, for
the transition between $\nu=1/3$ and the insulator: $p=1$; $g_1=g_0/3$ and $t_1=3g/g_0$; while for the transition
between $\nu=2/5$ and $\nu=1/3$: $p=2$; $g_2=(2/5-1/3)g_0$ and $t_2=\frac{g/g_{0}-1/3}{2/5-1/3}$, with $g$ being
the total conductance and $g_0=e^2/h$ the quantum conductance. The dependence of ithe charge on transmission, in
the simplest model, can be evaluated by considering the added current due to the two flux quanta attached to the
electron. Doing this, de Picciotto predicted \cite{raficm} the quasiparticle charge to vary from $e^*=e/(2p+1)$ at
$t_p\approx1$ to $e^*=e/(2p-1)$ at $t_p\approx0$ as a linear function of $t_p$, namely, for $p=1$ $e/3\rightarrow
e$ and for $p=2$ $e/5\rightarrow e/3$.
\begin{figure}[t]
\centering\epsfig{file=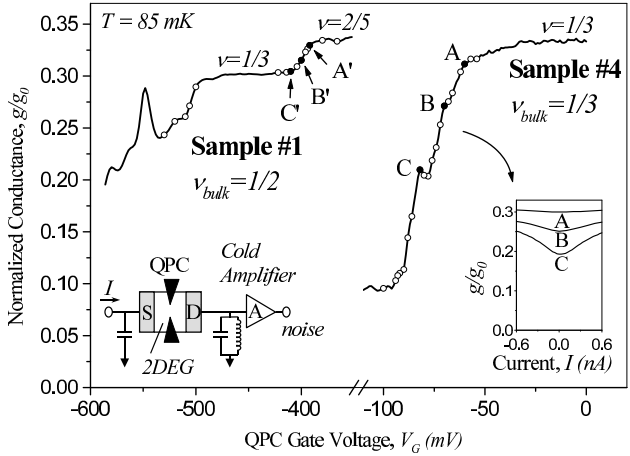,width=6cm,angle=270}\nopagebreak[3]
\vspace{0.5cm}\nopagebreak[3]\caption{Two-terminal conductance as a function of QPC gates voltage for samples
$\#1$ and $\#4$. The deviations from the quantized values of the conductance are due to the bulk longitudinal
resistance. The markers show the conductance values at which conductance and noise measurements were made. Right
Insert: Conductance as a function of applied DC current at the points shown. Left Insert: Schematic of sample and
measurement system.} \label{fig1}
\end{figure}%
In order to apply the above principles in a realistic experiment a more general expression for the shot noise
\cite{Martin92} applicable at finite temperatures, has to be used \cite{rafi97,Reznikov99}:
\begin{equation}\label{lesovik}
S_T=2e^*Vg_pt_p(1-t_p)\left[ \coth \left( \frac{e^*V}{2k_BT}\right) -\frac{%
2k_BT}{e^*V}\right] +4k_BTg  .
\end{equation}
This equation leads to a finite noise at zero applied voltage, $S=4k_BTg$ - the Johnson-Nyquist formula. When
$V>V_T\sim2k_BT/e^*$ the noise approaches the linear behavior predicted by Eqs. (\ref{schky}) and (\ref{limits}).
\\
\\
Measuring quasiparticle charge in the strong back-scattering limit is difficult, and results so far were
inconclusive \cite{Glattli99}.  As the QPC constriction is closed to reflect a larger portion of the incident
current, the conductance exhibits the familiar {\em impurity resonances} as a function of constriction width
(\cite{Milliken97}, and see also in Fig. 1).  Moreover, the $I-V$ characteristic becomes highly nonlinear ($g$ and
$t$ depend on current), making the analysis difficult.  Measuring a large number of samples across the full range
of the transmission coefficient in the first two CF channels, $\nu=1/3$ and $\nu=2/5$, we found relatively
resonant-free samples. Moreover, we extended Eq. (\ref{lesovik}) to cases of nonlinear $I-V$ characteristics
allowing also the charge to change with the transmission coefficient. Consequently, we have found a universal
behavior of the charge as a function of transmission in the $\nu=1/3$ channel, and qualitatively quite different
behavior for the charge in $\nu=2/5$ channel.
\begin{figure}[t]
\centering\epsfig{file=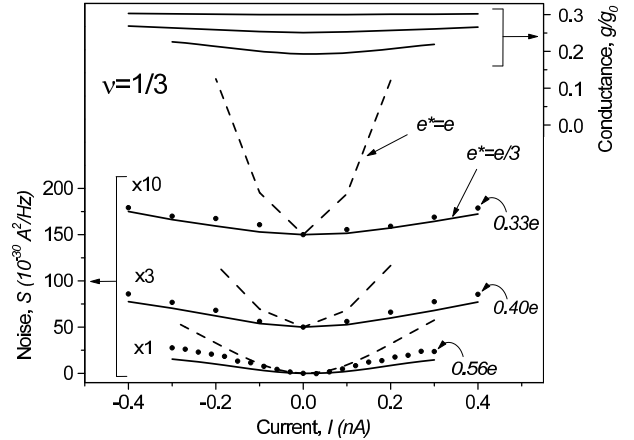,width=6cm,angle=270}\nopagebreak[3] \vspace{0.5cm}\nopagebreak[3]\caption{Top:
Differential conductance as a function of DC current for different transmissions in the $\nu=1/3$ channel for
sample $\#4$. Bottom: Measured excess noise as a function of DC current for the same transmission points. The
solid lines are the result of Eq. (\ref{lesmod}) with a charge $e^*=e/3$; the dashed lines are the result with
charge $e$. The numbers near the data points give the best-fit value to $e^*$ from Eq. (\ref{lesmod}).}
\label{fig2}
\end{figure}%
Our samples were 2DEG's embedded in GaAs-AlGaAs heterostructures with a low-temperature concentration of
$9.8\times10^{10}cm^{-2}$ and a mobility of $4\times10^6cm^2/Vs$.  A perpendicular magnetic field of $12.15T$ is
needed to reach the center of the $\nu=1/3$ plateau. The left-hand insert in Fig. 1 shows the schematic of the
two-terminal Hall samples with source (S), drain (D) and a QPC.  The Hall sample's width was $100\mu m$ and the
QPC opening width was $300nm$.  The QPC gate's potential was used to control the partitioning of the incoming
current.  Measurements were made in a dilution refrigerator at a lattice temperature of $55mK$ and a measured
electron temperature of $85mK$ (see \cite{rafi97} for details). Noise was measured within a bandwidth of $30kHz$
around a frequency of $1.6MHz$, chosen to be above the $1/f$-noise knee and much lower than $eV/h$.  An LRC
circuit determined the central frequency and bandwidth, with $R$ dominated by the resistance of the QPC and $C$ by
the capacitance of the coaxial lines.  A cold preamplifier, with a current noise of $\sim3\times10^{-29}A^2/Hz$,
amplified the noise signal.
\begin{figure}[t]
\centering\epsfig{file=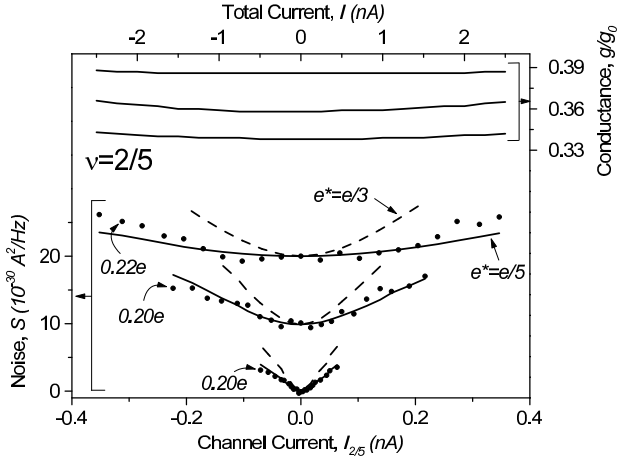,width=6cm,angle=270}\nopagebreak[3] \vspace{0.5cm}\nopagebreak[3]\caption{Top:
Differential conductance as a function of DC current for different transmissions in the $\nu=2/5$ channel for
sample $\#1$. Bottom: Measured excess noise as a function of DC current for the same transmission points. The
solid lines show the result of Eq. (\ref{lesmod}) with a charge $e^*=e/5$; the dashed lines are the result with
charge $e^*=e/3$. The expected noise with charge $e$ lies much above that of $e/3$.} \label{fig3}
\end{figure}%
We present here results from four samples $(\#1-\#4)$: three measured in the $\nu=1/3$ FQH state and two in the
$\nu=2/5$ FQH state. The bare samples (without applied gate voltage) exhibit, as a function of magnetic field, an
accurate $\nu=1/3$ quantization of the resistance but deviate at the $\nu=2/5$ plateau due to finite bulk
longitudinal resistance.  The measurements in the $\nu=2/5$ state were conducted at two different bulk filling
factors: $\nu_{bulk}=2/5$ and $\nu_{bulk}=1/2$ (see sample $\#1$ in Fig. 1), while for the measurements in the
$\nu=1/3$ state the bulk filling factors were $\nu_{bulk}=1/3$ and $\nu_{bulk}=1/2$ (see sample $\#4$ in Fig. 1).
Typical problems are seen in Fig. 1: sample $\#1$ shows a single large 'resonance' - the large spike on the
left-hand side of the graph - which prohibits further measurement into the $1/3$ state; and the reduction of the
transmission of the $1/3$ state in sample $\#4$, although much smoother, saturates at about $0.1e^2/h$, presumably
due to leakage across the QPC. The open circles on the graphs show where noise and $I-V$ measurements were made.
\\
\\
In our experiment we measured two quantities: the differential conductance $g$ and the shot noise. Using $g\propto
e^*t$ and $S_T$ from Eq. (\ref{lesovik}) we extracted the transmission probability $t$ and the quasiparticle's
charge $e^*$. However, the analysis is complicated by the strong dependence of the conductance on the current -
see the right-hand insert in Fig. 1. This insert shows the differential conductance of the QPC as a function of DC
current for three different conductances indicated by points A, B, and C. While at point A, where $t$ is
relatively large, the conductance is almost constant with current $(\Delta g/g_{I=0}=0.05)$, at point C, where $t$
is very small, there is a significant change in the differential conductance at large currents $(\Delta
g/g_{I=0}=0.3)$. To account for this non-linearity, the energy independent Eq. (\ref{lesovik}) was modified by
resorting to the integral over energy used in its derivation \cite{Martin92}. However, the dependence of
conductance on the current (in a small range), for a fixed QPC width, was all attributed to a changing $t$, i.e.,
the charge $e^*$ was approximated not to vary with current. Transforming from the integration over energy to a sum
over discreet current points, and substituting $t$ in terms of $g$ and $e^*$ in Eq. (\ref{lesovik}),
$t_{p=1}=\frac{(g_i/g_0)}{e^*/e}$, we get for $\nu=1/3$:
\begin{eqnarray}\label{lesmod}
S_{T}(I)=2e^{*}I\frac{1}{N}\sum_{i=1}^{N}\left( 1-\frac{g_{i}/g_{0}}{%
e^{*}/e}\right) \left[ \coth \left( \frac{e^{*}V}{2k_{B}T}\right)\right.&&\nonumber\\&&\hspace{-2.5cm}\left.-\frac{%
2k_{B}T}{e^{*}V}\right]+4k_{B}Tg  .
\end{eqnarray}
Here $i$ runs over the measured points $(N)$ up to current $I$ and $g_i$ is the differential conductance at each
point. In the $\nu=2/5$ state we substitute for the total current $I_T$ only that fraction which flows through the
2nd edge channel (using the CF model), $I_{p=2}=\frac{(g/g_0)-1/3}{g/g_0}I_T$, and for the transmission
$t_{p=2}=\frac{(g/g_0)-1/3}{(2/5-1/3)5e^*/e}$. Indeed, if $e^*=e/5$, $t_{p=2}$ is the expected bare transmission
of the 2nd CF channel given above. The noise expression now contains a single fitting parameter $e^*$.
\\
\\
Figure 2 shows noise results for a partitioned $\nu=1/3$ channel in sample $\#4$.  There is no noise on the
$\nu=1/3$ plateau.  The top part of the graph shows the differential conductance of the QPC against DC current $I$
at points A, B, and C shown in Fig. 1. The current range we used for the extraction of the charge is $\Delta
g/g_{I=0}=0-0.2$ in order to reduce the effect of the charge variation with current while still being able to fit
the curves to Eq. (\ref{lesmod}). The measured noise, with the background thermal noise subtracted, is shown in
the lower part of Fig. 2. The curves are offset for clarity. Also shown is the behavior of Eq. (\ref{lesmod}) with
$e^*=e/3$ (solid lines) and $e^*=e$ (dashed lines). For each width of the QPC constriction we find the best
fitting quasiparticle charge $e^*$ and consequently the channel transmission $t$ near $I=0$. In previously
published high-$t$ data the noise is that of $e/3$ charges \cite{rafi97}. As the transmission is reduced the
apparent charge increases to a maximum around charge $e$. Consistent results were obtained for the two other
samples (as seen in Fig. 4). Similarly, Fig. 3 shows similar graphs for the measurements in the $\nu=2/5$ state in
sample $\#1$ (points A', B', and C' in Fig. 1). Again, no noise is measured on the $\nu=2/5$ plateau. The
theoretical lines correspond to charges $e^*=e/5$ (solid lines) and $e^*=e/3$ (dashed lines). The other sample
provided similar results.
\begin{figure}[t]
\centering\epsfig{file=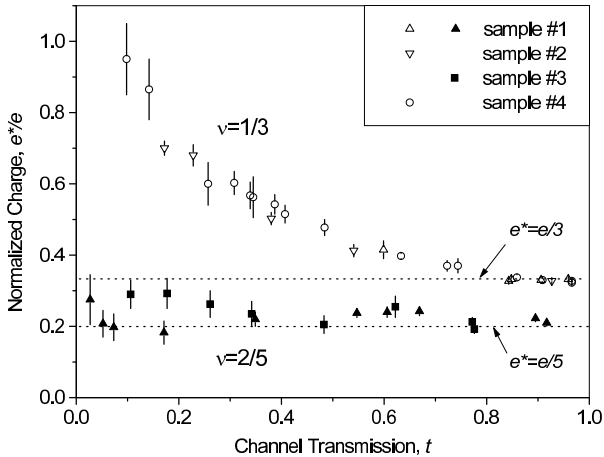,width=6cm,angle=270}\nopagebreak[3]
\vspace{0.5cm}\nopagebreak[3]\caption{Summary of the results of the determined evolution of the charge of the
quasiparticles as a function of transmission, for all four samples, for the $\nu=1/3$ and $\nu=2/5$ channels.}
\label{fig4}
\end{figure}
The dependences of the quasiparticle charge on transmission coefficient for all four samples are summarized in
Fig. 4.  All results approximately collapse onto two separate curves. While in the $\nu=1/3$ case the deduced
charge changes smoothly from $e/3$ at weak reflection (large $t$) to around $e$ at strong reflection
($t\cong0.1$), the deduced charge in the $\nu=2/5$ case stays near $e/5$ over almost the full range of
transmission. There is an apparent slight increase of $e^*$ at lower transmissions. Although scattering of the
data due to the small signal prevents a more accurate determination of the charge for $t<0.3$, it clearly does not
show the steep rise to $e^*=e$ observed at $\nu=1/3$.
\\
\\
Adopting the CF picture in accordance with Ref. 11, the difference between the two channels can be understood by
considering how much charge crosses the constriction when a composite fermion, composed of an electron and two
flux quanta, traverses it. In the $\nu=1/3$ case, a strongly closed constriction, reflecting almost all the
incident current, is almost an insulator and the extra charge induced by the fluxes is negligible, leading to a
quasiparticle charge approximately $e$. In contrast, in the $\nu=2/5$ case only one of the edge channels is
strongly reflected, and consequently the constriction is not an insulator. Thus the extra transferred charge is
finite and the quasiparticle's charge is not $e$. Eqs. 1-4 are based on a picture in which the noise is produced
by independent quasiparticles whose partitioning obeys binomial statistics. In fact the noise can be interpreted
also as being generated by quasiparticles of fixed charge whose partitioning statistics are not binomial. For
example, the measured charge of $e^*=e$ could be interpreted as a quasiparticle of charge $e$ (a single electron)
or as three quasiparticles of charge $e^*=e/3$ bunched together. For the $\nu=2/5$ channel, we may conclude that
the $e^*=e/5$ quasiparticles traverse an opaque barrier without fully bunching, which would produce a charge
$e^*=e$. As yet, there is no rigorous theory for the $\nu=2/5$ case.
\\
\\
We thank M.~Reznikov and R.~de~Picciotto for useful discussions and guidance. We also thank F.~von~Oppen for
instructive discussions. The work was partly supported by the Israeli Academy of Science, the Israel-USA
Binational Science Foundation and the Israel-Germany DIP grant.

\end{document}